# INCREASE OF AMPLITUDE OF ACCELERATING WAKEFIELD EXCITED BY SEQUENCE OF SHORT RELATIVISTIC ELECTRON BUNCHES IN PLASMA AT MAGNETIC FIELD USE

**D.S. Bondar[1], I.P. Levchuk[2], V.I. Maslov[2], I.N. Onishchenko[2]**

[1] *Karazin Kharkiv National University*
*61022, Kharkiv, Ukraine*
[2] *NSC Kharkov Institute of Physics & Technology*
*61108 Kharkiv, Ukraine*
*e-mail: vmaslov@kipt.kharkov.ua*

Earlier, the authors found a mechanism for the sequence of short relativistic electron bunches, which leads to resonant excitation of the wakefield, even if the repetition frequency of bunches differs from the plasma frequency. In this case, the synchronization of frequencies is restored due to defocusing of the bunches which get into the bad phases with respect to the plasma wave. However, in this case, the bunches are lost, which as a result of this do not participate in the excitation of the wakefield. In this paper, numerical simulation was used to study the dynamics of electron bunches and the excitation of the wakefield in a magnetized plasma by a long sequence of short bunches of relativistic electrons. When a magnetic field is used, the defocussed bunches return to the region of interaction with the field after a certain time. In this case, the electrons of the bunches, returning to the necessary phases of the field, participate in the excitation of the wakefield. Also, the use of a magnetic field leads to an increase of the frequency of the excited wave relative to the repetition frequency of bunches. The latter increases the time for maintaining the resonance and, consequently, leads to an increase of the amplitude of the excited wakefield.

**KEYWORDS:** wakefield, relativistic electrons, maintenance of the existence of a resonance, an increase of the amplitude of the excited wakefield.

## ЗБІЛЬШЕННЯ АМПЛІТУДИ ПРИСКОРЮЮЧОГО КІЛЬВАТЕРНОГО ПОЛЯ, ЯКЕ ЗБУДЖУЄТЬСЯ ПОСЛІДОВНІСТЮ КОРОТКИХ РЕЛЯТИВІСТСЬКИХ ЕЛЕКТРОННИХ ЗГУСТКІВ В ПЛАЗМІ, ПРИ ВИКОРИСТАННІ МАГНІТНОГО ПОЛЯ

**Д.С. Бондар[1], І.П. Левчук[2], В.І. Маслов[2], І.М. Онищенко[2]**

[1] *Харківський національний університет імені В.Н. Каразіна*
*61022, Харків, Україна*
[2] *ННЦ Харківський фізико-технічний інститут*
*61108, Харків, Україна*

Раніше авторами був знайдений для послідовності коротких релятивістських електронних згустків механізм, який призводить до резонансного збудження кільватерного поля, навіть якщо частота проходження згустків відрізняється від плазмової частоти. В цьому випадку синхронізація частот відновлюється за рахунок дефокусування згустків, які потрапляють в погані фази по відношенню до плазмової хвилі. Однак при цьому втрачаються згустки, які в результаті цього не беруть участі в збудженні кільватерного поля. У цій роботі чисельним моделюванням вивчена динаміка електронних згустків і збудження кільватерного поля в замагніченій плазмі довгою послідовністю коротких згустків релятивістських електронів. При використанні магнітного поля дефокусовані згустки через певний час повертаються в область взаємодії з полем. При цьому електрони згустків, що повертаються в потрібні фази поля, беруть участь в збудженні кільватерного поля. Також використання магнітного поля призводить до збільшення частоти збуджуваної хвилі щодо частоти проходження згустків. Останнє збільшує час підтримки резонансу і, отже, призводить до збільшення амплітуди кільватерного поля.

**КЛЮЧОВІ СЛОВА:** кільватерне поле, релятивістські електрони, підтримання існування резонансу, збільшення амплітуди збуджуваного кільватерного поля.

## УВЕЛИЧЕНИЕ АМПЛИТУДЫ УСКОРЯЮЩЕГО КИЛЬВАТЕРНОГО ПОЛЯ, ВОЗБУЖДАЕМОГО ПОСЛЕДОВАТЕЛЬНОСТЬЮ КОРОТКИХ РЕЛЯТИВИСТСКИХ ЭЛЕКТРОННЫХ СГУСТКОВ В ПЛАЗМЕ, ПРИ ИСПОЛЬЗОВАНИИ МАГНИТНОГО ПОЛЯ

**Д.С. Бондарь[1], И.П. Левчук[2], В.И. Маслов[2], И.Н. Онищенко[2]**

[1] *Харьковский национальный университет имени В. Н. Каразина*
*61022, Харьков, Украина*
[2] *ННЦ Харьковский физико-технический институт*
*61108, Харьков, Украина*

Ранее авторами был найден для последовательности коротких релятивистских электронных сгустков механизм, который приводит к резонансному возбуждению кильватерного поля, даже если частота следования сгустков отличается от плазменной частоты. В этом случае синхронизация частот восстанавливается за счет дефокусировки сгустков, которые попадают в плохие фазы по отношению к плазменной волне. Однако при этом теряются сгустки, которые в результате этого не участвуют в возбуждении кильватерного поля. В этой работе численным моделированием изучена динамика электронных сгустков и возбуждение кильватерного поля в замагниченной плазме длинной последовательностью коротких сгустков релятивистских электронов. При использовании магнитного поля дефокусированные сгустки через определенное время возвращаются в область взаимодействия с полем. При этом электроны сгустков, возвращающиеся в нужные фазы поля, участвуют в возбуждении кильватерного поля. Также использование магнитного поля приводит к увеличению частоты возбуждаемой волны относительно частоты следования сгустков. Последнее увеличивает время поддержания резонанса и, следовательно, приводит к увеличению амплитуды возбуждаемого кильватерного поля.

**КЛЮЧЕВЫЕ СЛОВА***: кильватерное поле, релятивистские электроны, поддержание существования резонанса, увеличению амплитуды возбуждаемого кильватерного поля*

Development of accelerators of charged particles, in particular, colliders is one of the most promising areas of current research. Modern accelerators on metal structures are huge and expensive, because metal breaks out at 100MeV/m. It is clear that to achieve in linear colliders high energy (above 1 TeV) it is necessary to increase their length to tens of kilometers. One can estimate which dimensions of the accelerators should be used to accelerate electrons to the required energy of 1 TeV. It's tens of kilometers. It is necessary to make them smaller and cheaper. To do this, it is necessary to increase the rate of acceleration, i.e. to increase the accelerating field. In a plasma, one can excite the field $E_z$=100 GeV/m. This can be done using a bunch of electrons, by a bunch of ions or by a laser pulse. In the experiment, the record results were already obtained: the laser pulse accelerated electrons in the plasma to 4.2 GeV at a distance 9cm. Therefore, the electric field equals $E_z \approx$47 GeV/m [1]. Also in an experiment in plasma, a dense electron bunch with the energy of 42 GeV excited the wakefield, and its tail accelerated to the energy 84 GeV (i.e., doubled the energy) at a distance of approximately 1 m [2], then the electric field is $E_z \approx$42 GeV/m. In a dielectric in the experiment, an impulse field of approximately 10 GeV/m has been excited. I.e. the dielectric accelerator can be in 100 times shorter than the metallic accelerator, and the plasma accelerator is in 1000 times shorter. Because the dielectric accelerator is easier to operate, and the plasma provides larger fields, the dielectric and plasma accelerators are intensively investigated.

The wakefield excitation in a plasma by a long sequence of electron bunches is considered in this paper. An important factor of interest to this case is the use of external longitudinal (along the axis) magnetic field with the purpose to increase the accelerating field. So, the aim of the work is to investigate the features of use some optimal magnetic field, that would ensure a greater growth of the rate of acceleration.

## PROBLEM STATEMENT

At resonant excitation of the wakefield in the plasma, the repetition frequency of bunches $\omega_m$, is equal to the frequency of the excited wakefield $\omega_{pe}=\omega_m$, i.e. to electron plasma frequency $\omega_{pe}$. Since along the radius r, the wakefield is localized near the sequence of bunches, i.e. along the radius r, it is localized in some neighborhood of the sequence of bunches, then the wakefield has not only a longitudinal $E_z$ accelerating/decelerating field, but also a radial $F_r$ focusing/defocusing force. The field $E_z$ and $F_r$ are shifted relative to each other by a quarter of the wavelength, i.e. on $\pi/2$. Where $E_z$=$E_{z\,max}$, there $F_r$=0; and where $E_z$=0, there $F_r$=$F_{r\,max}$. Since the bunches are finite size, the part of the bunch is defocused and ceases to excite wakefield. Part firstly is focused, and then due to the expansion of betatron oscillations (i.e., radial in the radial potential well) is again defocused, i.e. it leaves along r the region of interaction with the field and ceases to excite the wakefield too. Therefore, it is advisable not to allow the bunches to be defocused or periodically return them by an external longitudinal (along the axis) magnetic field. Although it is known that the use of a magnetic field in the experiment is associated with additional difficulties. But the magnetic field suppresses the focusing, and the maximum wakefield has been observed when the bunches are focused by the wakefield. Also, the magnetic field suppresses defocusing. And it is very important for the excitation of the wakefield, because in the experiment it is very difficult to maintain the resonant plasma because of its uncontrolled inhomogeneity and nonstationary. And in the nonresonant case only small amplitude beatings are excited. However, it was shown in [3-7] that due to the self-cleaning of the "bad" bunches at their defocusing and in the nonresonant case, intense excitation of the wakefield is possible.

## ANALYSIS OF THE USING OF A MAGNETIC FIELD

As a consequence, it would seem that the use of a magnetic field is impractical, it suppresses defocusing. Hence it follows that one must use some optimal magnetic field when it does not yet suppress focusing and defocusing, but it already returns defocussed bunches after some time into the region of interaction with the field. Then it is necessary to use a not very optimal magnetic field, so that after defocusing the bunches they return to the axis after some time, and they again excite the wakefield. The return time is $2\pi/\omega_{ce}$. Where $\omega_{ce}$ is the electron cyclotron frequency. In order that $H_0$ allows electrons to reach the axis in the focusing field, the radius of the radial oscillations of electrons in crossed $H_{0z}$ and $E_r$ fields should be not less than the radius of the bunch

$$\frac{eE_r}{m_e \gamma_b \omega_{ce}^2} > r_b. \tag{1}$$

It is necessary that for N-th bunch (if N bunches excite the maximum wakefield) it is satisfied when the maximum wakefield is reached. Sequential bunches excite wakefield that firstly it grows linearly with increasing number of injected bunches

$$E_{rN} = NE_{r1} . \tag{2}$$

$E_{rN}$ is the radial wakefield after N bunches. $E_{r1}$ is the radial wakefield after the 1st bunch. Then it must be fulfilled

$$\frac{eNE_{r1}}{m_e \gamma_b \omega_{ce}^2} > r_b . \tag{3}$$

It is necessary to note that an additional frequency shift occurs when a magnetic field is used which, at a certain value of the magnetic field, tightens the maintenance of the resonance, leading to an increase of the excited wakefield.

A long sequence of electron bunches of a small charge was used in the experiment [8] and in numerical simulation [9] for excitation of an intense wakefield in a plasma. However, as it turned out, there is the limiting amplitude of the wakefield. It is determined by the fact that the nonlinear shift of the frequency of the wakefield appears with increase of the wakefield amplitude. Because of this, the resonant interaction of bunches with the wakefield is detuned. This resonance detuning is delayed in time, if, as it was done in [10], a small excess of the plasma density above the resonant value is chosen initially. The same resonance maintenance can be ensured by using a small magnetic field. Using the LCODE [11], numerical simulation of the growth of the wakefield amplitude was performed. It is shown that when the resonance is maintained, the amplitude of the wakefield increases in comparison with the case of the initial resonant conditions.

The wakefield excitation in a plasma by a long sequence of electron bunches is considered in this paper.

The resonant excitation of the wakefield by a long sequence of relativistic electron bunches is difficult, because it is difficult to maintain a homogeneous and stationary plasma in the experiment. However, intense wakefield excitation by a long sequence of relativistic electron bunches has been observed. The mechanism of resonant excitation of the wakefield by a nonresonant sequence of short electron bunches has been investigated in [9]. Frequency synchronization is carried out due to self-cleaning of the sequence of bunches due to defocusing and leaving along radius of some bunches that are not in phase with the wave. The simulation results of the mechanism of maintaining the resonance of electron bunches with a wakefield using a magnetic field are presented in this paper.

In [9], for a sequence of short relativistic electron bunches, a mechanism was found which leads to a resonant excitation of the wakefield, even if the repetition frequency of bunches differs appreciably from the plasma frequency. Synchronization of frequencies is restored due to defocusing of bunches which get into bad phases with respect to the plasma wave. However, the bunches are lost, which as a result do not participate in the excitation of the wakefield. When a magnetic field is used, the defocused bunches return to the region of interaction with the field after a certain time. In this case, the electrons of the bunches returning to the necessary phases of the field can participate in the excitation of the wakefield. Also, the use of a magnetic field leads to an increase of the frequency of the excited wave relative to the repetition frequency of bunches.

The latter increases the time for maintaining the resonance and, consequently, leads to an increase of the amplitude of the excited wakefield.

## SIMULATION OF RESONANCE RECOVERY FOR THE CASE OF 32 BUNCHES

For numerical simulation parameters are selected: $n_{res} = 10^{11} cm^{-3}$ is the resonant plasma density which corresponds to ratio $\omega_{pe} = \omega_m = 2\pi \cdot 2.8 \cdot 10^9$, relativistic factor of bunches equals $\gamma_b = 5$, have been selected. Where $\omega_m$ is the repetition frequency of bunches, $\omega_{pe} = (4\pi n_{res} e^2/m_e)^{1/2}$ is the electron plasma frequency. The density of bunches $n_b = 6 \times 10^8 cm^{-3}$ is distributed in the transverse direction approximately according to Gaussian distribution, $\sigma_r = 0.5 cm$, $\lambda = 10.55 cm$ is the wavelength, $\xi = V_b t - z$, $V_b$ is the velocity of bunches. Time is normalized on $\omega_{pe}^{-1}$, distance - on $c/\omega_{pe}$, density - on $n_{res}$, current $I_b$ - on $I_{cr} = \pi mc^3/4e$, fields – on $(4\pi n_{res} c^2 m_e)^{1/2}$.

We consider the dynamics of the first 32 bunches in plasma. We use the cylindrical coordinate system (r, z) and draw the plasma and beam densities at some z as a function of the dimensionless time $\tau = \omega_p t$.

The longitudinal coordinate $\xi = z - V_b t$ is normalized on $2\pi/\lambda$ ($\lambda$ is the wavelength). The values of the $E_z$, $F_r$, $H_\theta$ and $H_0$ are normalized on $mc\omega_{pe}/e$. Where e, m are the charge and mass of the electron, c is the light velocity, $\omega_{pe}$ is the electron plasma frequency.

We do not take into account the longitudinal dynamics of the bunches, because at the times and energies of the beam according to

$$\frac{dV_z(r)}{dr} \propto \frac{1}{\gamma_b^3} , \quad \frac{dV_r(r)}{dr} \propto \frac{1}{\gamma_b}$$

radial relative shifts of beam particles predominate. $V_z$, $V_r$ are the longitudinal and radial velocities of the electron bunches, $\gamma_b$ is the relativistic factor of the bunches.

The wakefield excitation by 32 bunches is considered for two cases: initially the resonant case and the case of resonance recovery. We consider a sequence of 32 bunches which are uniform in the longitudinal direction and are distributed according to Gaussian along the radius. The lengths of the bunches are chosen equal to half of the wavelength $\xi_b = \lambda/2$.

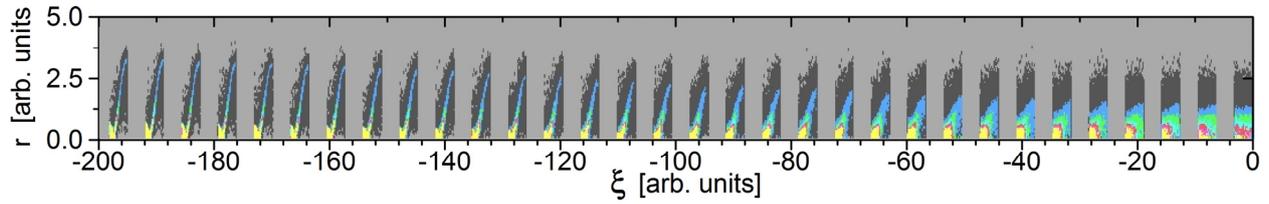

Fig. 1. Spatial distribution of density $n_b$ of sequence of initially radially Gaussian and longitudinally homogeneous resonant bunches into the plasma at $\xi_b = \lambda/2$, $I_b = 0.2 \times 10^{-3}$

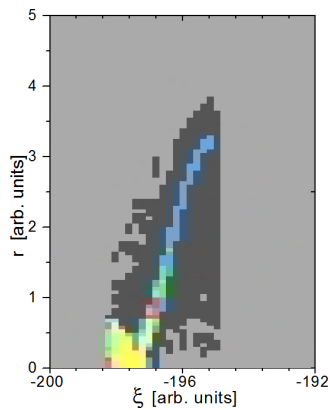

Fig. 1a (part of Fig. 1). Spatial distribution of density $n_b$ of 32nd initially radially Gaussian and longitudinally homogeneous resonant bunch into the plasma at $\xi_b = \lambda/2$, $I_b = 0.2 \times 10^{-3}$

We consider the case when the initial density of plasma electrons $n_0$ is such that the repetition frequency of the bunches $\omega_m$ is equal to the electron plasma frequency $\omega_m = \omega_{pe}$. In this case, up to the point of maximum focusing of the bunches $E_z$ grows and then it decreases during refocusing. To compensate the charge of the bunches, some of the plasma electrons leave the axis. Then the phase velocity of the wave $V_{ph}(r=0)$ (and $\omega_{pe}(r=0)$) on the axis becomes smaller than at the periphery with respect to r. As a result, the wave becomes skewed (Fig. 2.).

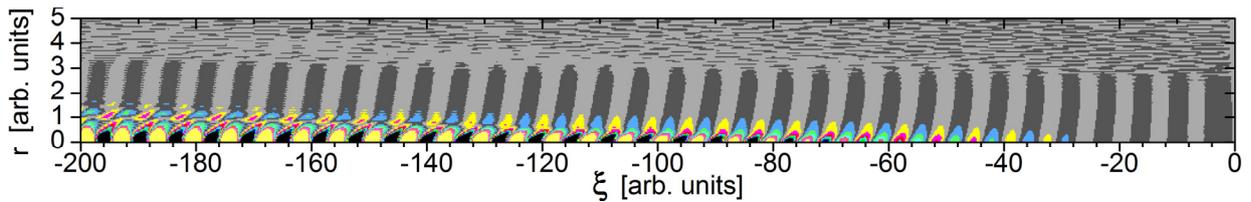

Fig. 2. Spatial distribution of plasma electron density $n_e$ in wakefield, excited by sequence of initially radially Gaussian and longitudinally homogeneous resonant bunches

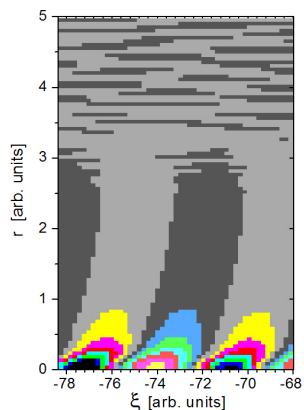

Fig. 2a (part of Fig. 2). Spatial distribution of plasma electron density $n_e$ in 12th wavelength of wakefield, excited by sequence of initially radially Gaussian and longitudinally homogeneous resonant bunches

Also, in this case, $V_{ph}(r=0) < V_b$ is smaller than the beam velocity $V_b$, and $\omega_{pe}(r=0) < \omega_m$, i.e. the wave becomes nonresonant with a sequence of bunches. The wave lags behind the bunches and the smaller their parts get into the focusing fields (Figs. 1, 1a, 3, 3a, 5, 5a, 7, 7a).

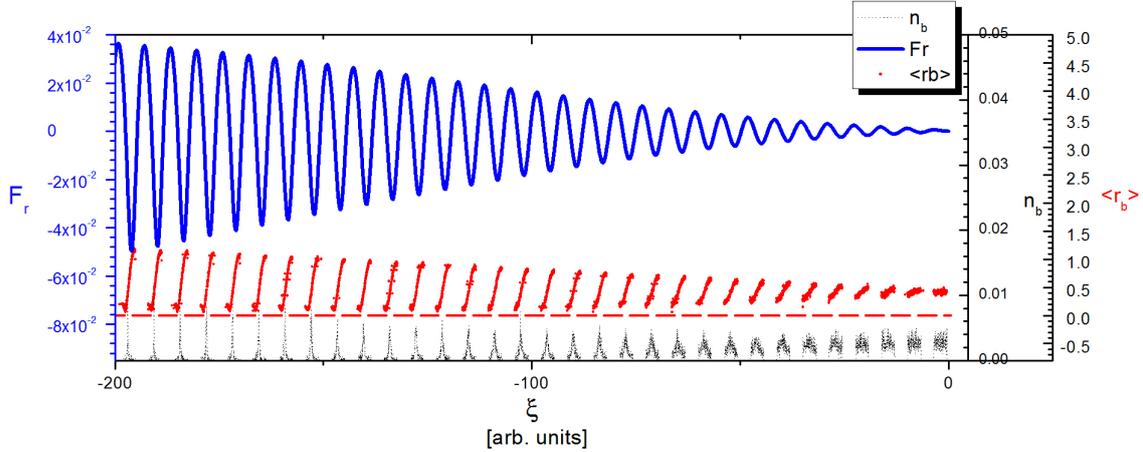

Fig. 3. Longitudinal distribution of radius $r_b$, of density $n_b$ of sequence of initially radially Gaussian and longitudinally homogeneous resonant bunches and of radial wake force $F_r$ into the plasma at $\xi_b = \lambda/2$, $I_b = 0.2 \times 10^{-3}$

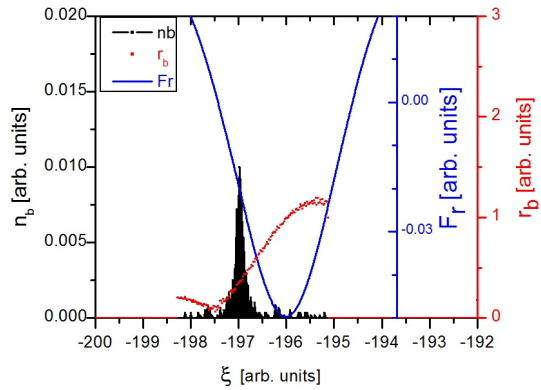

Fig. 3a (part of Fig. 3). Longitudinal distribution of radius $r_b$, of density $n_b$ of 32nd initially radially Gaussian and longitudinally homogeneous resonant bunch and of radial wake force $F_r$ in 32nd wavelength of wakefield into the plasma at $\xi_b = \lambda/2$, $I_b = 0.2 \times 10^{-3}$

In this case, a very small part of their first front gets into the accelerating phases (Figs. 4, 4a.).

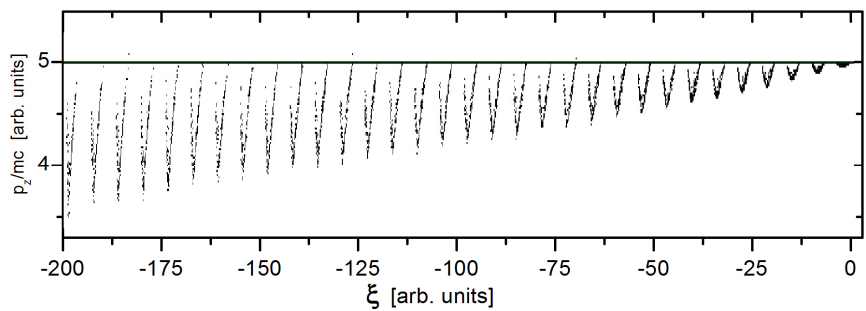

Fig. 4. Longitudinal momenta of 32 bunches

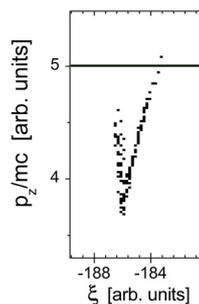

Fig. 4a (part of Fig. 4). Longitudinal momenta of 30th bunch

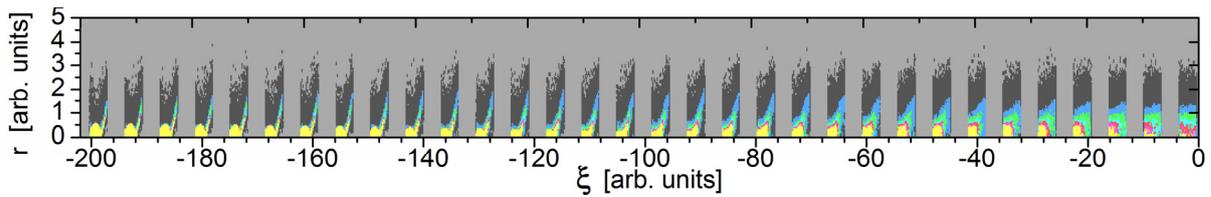

Fig. 5. Spatial distribution of density $n_b$ of sequence of initially radially Gaussian and longitudinally homogeneous bunches into the plasma at $\xi_b=\lambda/2$, $I_b=0.2\times10^{-3}$, $H_0=0.1$, $n_e/n_{res}-1=0.02$, $n_{res}$ is the resonant plasma electron density

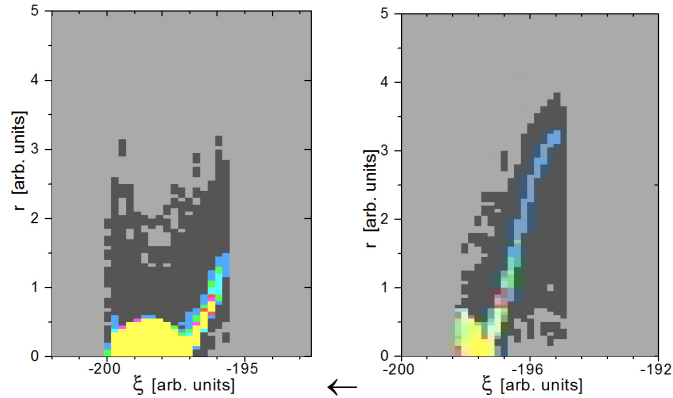

Fig. 5a (part of Fig. 5). Spatial distribution of density $n_b$ of 32nd initially radially Gaussian and longitudinally homogeneous resonant (right) (left at $H_0=0.1$, $n_e/n_{res}-1=0.02$) bunch into the plasma at $\xi_b=\lambda/2$, $I_b=0.2\times10^{-3}$

If we use a somewhat larger $n_0$ and a small magnetic field $H_0=0.1$, then the resonance of the wave with the bunch sequence is restored on the axis, since on the axis $\omega_{pe}(0)\approx\omega_m$. However, at the periphery with respect to $r$, the wave becomes nonresonant with the bunch sequence, so now $\omega_{pe}(r>0)>\omega_m$. Then the wave becomes skewed in the opposite direction (Figs. 2, 2a, 6, 6a.).

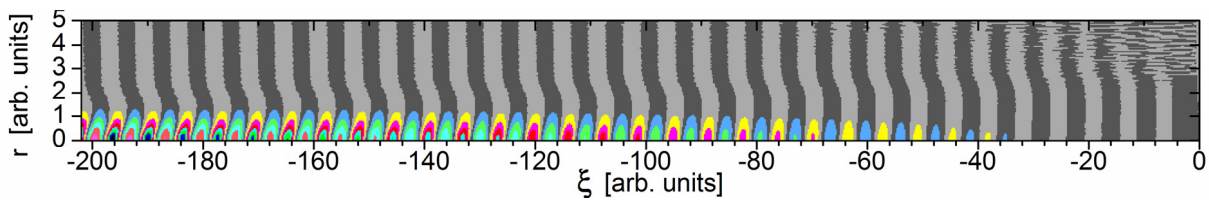

Fig. 6. Spatial distribution of plasma electron density $n_e$ in wakefield, excited by sequence of initially radially Gaussian and longitudinally homogeneous bunches into the plasma at $\xi_b=\lambda/2$, $I_b=0.2\times10^{-3}$, $H_0=0.1$, $n_e/n_{res}-1=0.02$

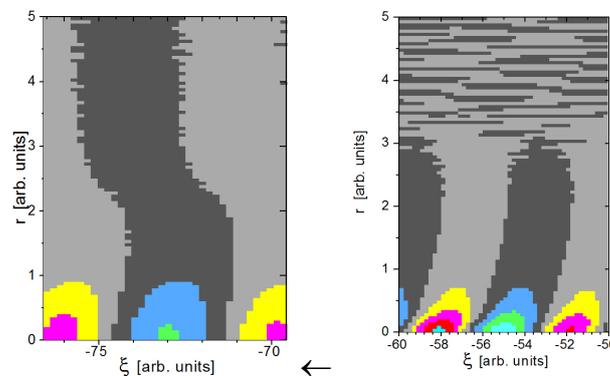

Fig. 6a (part of Fig. 6). Spatial distribution of plasma electron density $n_e$ in wakefield, excited by sequence of initially radially Gaussian and longitudinally homogeneous resonant (rigth) (left at $H_0=0.1$, $n_e/n_{res}-1=0.02$) bunch into the plasma at $\xi_b=\lambda/2$, $I_b=0.2\times10^{-3}$, $H_0=0.1$, $n_e/n_{res}-1=0.02$

When the resonance of a wave is restored to a sequence of bunches on the axis, larger parts of the bunches get into the focusing fields (Figs. 1, 1a, 3, 3a, 5, 5a, 7, 7a), and defocused bunches is weakly defocused (Fig. 7a),

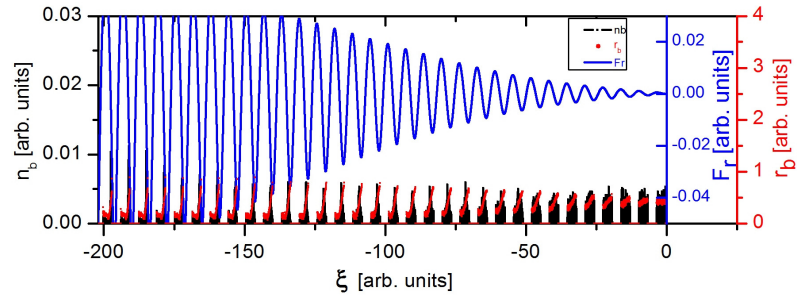

Fig. 7. Longitudinal distribution of radius $r_b$ of density $n_b$ of sequence of initially radially Gaussian and longitudinally homogeneous bunches and of radial wake force $F_r$ into the plasma at $\xi_b = \lambda/2$, $I_b = 0.2 \times 10^{-3}$, $H_0 = 0.1$, $n_c/n_{res} - 1 = 0.02$

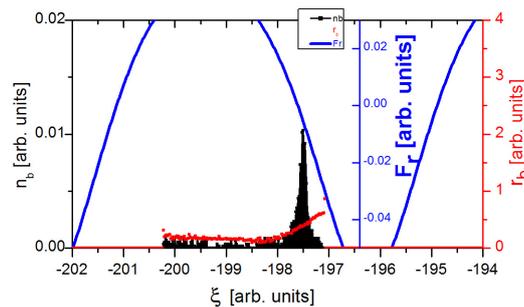

Fig. 7a (part of Fig. 7). Longitudinal distribution of radius $r_b$ of density $n_b$ of bunch and of radial wake force $F_r$ into the plasma at $H_0 = 0.1$, $n_c/n_{res} - 1 = 0.02$) the excited wakefield becomes larger (Figs. 3,7) compared with the initially resonant and subsequent disturbance of the resonance, and small parts of the back fronts of the bunches get into the accelerating phases (Figs. 8).

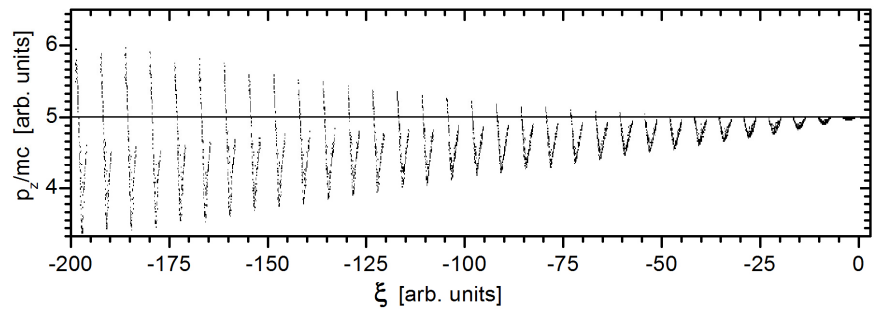

Fig. 8. Longitudinal momenta of 32 bunches at $\xi_b = \lambda/2$, $I_b = 0.2 \times 10^{-3}$, $H_0 = 0.1$, $n_c/n_{res} - 1 = 0.02$

At $H_0 = 0.1$ the wakefield $E_z$ became more uniform along the plasma, because it does not so rapidly decrease to the end of the plasma.

## CONCLUSION

The main conclusion is that, magnetic field can be used for wakefield increase and for increase the transformation of driver-bunch energy into the accelerated electrons (witness) energy. In this paper it is shown that in order to improve the energy transformation of driver-bunch energy into the witness energy, some optimal magnetic field should be used when it does not yet suppress focusing and defocusing. Such optimal magnetic field should also ensure returns defocussed bunches after some time into the region of the interaction with the field. Moreover, it is important to choose such optimal magnetic field that would return the bunches to the axis, after some time that bunches would again excite the wakefield.

It is also important to note that the use of a magnetic field leads to an increase in the frequency of the excited wave relative to the repetition frequency of bunches. Using the code lcode [9], numerical simulation of the growth of the wakefield amplitude was performed. It is shown that when the resonance is maintained, the amplitude of the wakefield increases in comparison with the case of the initial resonant conditions. The simulation results of the mechanism of maintaining the resonance of electron bunches with a wakefield using a magnetic field are presented in this paper.